\documentclass{article}

\PassOptionsToPackage{numbers, sort, compress}{natbib}

\usepackage[final]{neurips_2022_ml4ps}

\usepackage[utf8]{inputenc} 
\usepackage[T1]{fontenc}    
\usepackage{hyperref}       
\usepackage{url}            
\usepackage{booktabs}       
\usepackage{amsfonts}       
\usepackage{nicefrac}       
\usepackage{microtype}      
\usepackage{xcolor}         

\usepackage{mathtools}
\usepackage{xspace}
\usepackage{amsmath}
\usepackage{bbm}
\usepackage{amssymb}
\usepackage{pifont}
\usepackage{caption}
\usepackage{subcaption}
\usepackage{graphicx}
\usepackage{wrapfig}
\usepackage{enumitem}
\usepackage{multirow}
\usepackage{comment}
\usepackage[acronym]{glossaries}
\usepackage{cleveref}

\newcommand{\bx}{\boldsymbol{x}}
\newcommand{\bs}{\boldsymbol{s}}

\newcommand{\bvartheta}{\boldsymbol{\vartheta}}

\newacronym{dm}{DM}{dark matter}
\newacronym{gce}{GCE}{galactic center excess}
\newacronym{ps}{PS}{point source}
\newacronym{lat}{LAT}{Large Area Telescope}
\newacronym{nre}{NRE}{neural ratio estimation}
\newacronym{mnre}{MNRE}{marginal neural ratio estimation}
\newacronym{tmnre}{TMNRE}{truncated marginal neural ratio estimation}
\newacronym{sbi}{SBI}{simulation-based inference}
\newacronym{bce}{BCE}{binary cross-entropy}
\newacronym{nn}{NN}{neural network}
\newacronym{mlp}{MLP}{multi-layer perceptron}
\newacronym{psf}{PSF}{point-spread function}
\newcommand{\swyft}{\textit{swyft}}

\title{Detection is truncation: studying source populations with truncated marginal neural ratio estimation}

\author{
	Noemi Anau Montel \\
	GRAPPA Institute \\ University of Amsterdam, The Netherlands \\ 
	\texttt{n.anaumontel@uva.nl}
	\And
	Christoph Weniger \\
	GRAPPA Institute\\ University of Amsterdam, The Netherlands \\
	\texttt{c.weniger@uva.nl}
}

\begin{document}
\maketitle

\begin{abstract}
  Statistical inference of population parameters of astrophysical sources is challenging. It requires accounting for selection effects, which stem from the artificial separation between bright detected and dim undetected sources that is introduced by the analysis pipeline itself.
  %
  We show that these effects can be modeled self-consistently in the context of sequential simulation-based inference. Our approach couples source detection and catalog-based inference in a principled framework that derives from the 
  truncated marginal neural ratio estimation (TMNRE) algorithm.  It relies on the realization that detection can be interpreted as prior truncation. We outline the algorithm, and show first promising results.
  %
\end{abstract}

\section{Introduction}
\label{sec:intro}

Point sources detection is crucial for astronomical surveys, and is the cornerstone for the compilation of source catalogues. Those source catalogues are then typically the basis for the inference of physical parameters that describe the sources at the population level.
Upcoming astronomical facilities, such as the Square Kilometer Array (SKA)~\citep{ska} and the Cherenkov Telescope Array (CTA)~\citep{cta} will deliver large and complex datasets.  In order to leverage their full potential, it is urgent to develop robust and automated source detection and source population parameters inference algorithms.

Recent developments in deep learning and more generally automatic differentiation frameworks~\citep{Baydin_2015} are increasingly used for tackling difficult astronomical data analysis challenges.
%
The capability of deep learning techniques of point sources detection and population characterization has been demonstrated across different wavelengths surveys, e.g. in $\gamma$-ray data~\citep{Panes_2021, Caron_2018, List_2021, Mishra-Sharma_2022}, radio data~\citep{VafaeiSadr_2018, Rezaei_2021, Lukic_2019, Tilley_2021} and cosmic microwave background data~\citep{Bonavera_2021}.
In particular simulation-based machine learning approaches can be highly flexible, allowing to tailor developed pipelines to specific telescopes and science cases. 
A range of \gls*{sbi} algorithms have been proposed in the literature (see \citet{Cranmer_2020} for a review). An appealing feature is that they generally allow to directly estimate marginal posteriors for parameters of interest~\citep{miller_2020}. Furthermore, \textit{sequential} \gls*{sbi} approaches~\citep{Durkan_2018, Papamakarios_2016, Papamakarios_2018} have been shown to be particularly simulation efficient.  Among those, \gls*{tmnre} \citep{miller_2020, miller_2021} is a sequential SBI approach 
based on \gls*{nre}~\citep{Hermans_2019},  which particularly well composes with marginalization.

Here, we present a strategy for how to use \gls*{tmnre} \footnote{
    \ We use \swyft \ \gls*{tmnre} implementation that can be found at \url{https://github.com/undark-lab/swyft}.
    }
to simultaneously perform source detection and population-level parameters inference. This enables to self-consistently combine information from both detected and sub-threshold sources, without being affected by detection biases. The key idea is to recast the traditional concept of \emph{source detection} in terms of \emph{prior truncation}.  
This will allow us to \textit{distill} information of bright sources directly into the simulation model.
Our proposed method is highly interpretable since it resembles components of traditional survey analysis workflows.

\section{Methodology}
\label{sec:method}


\paragraph{Background}

\gls*{tmnre} (and \gls*{nre} in general) performs posterior estimation by solving a binary classification problem.
%
Given a model $p(x, z) = p(x|z) p(z)$, where $x$ is data, and $z$ a set of parameters of interest, one trains a network to distinguish joined samples $x, z \sim p(x, z)$ from marginal samples $x, z\sim p(x) p(z)$.  The networks learns to estimate the likelihood-to-evidence ratio $r(z; x)=p(x|z)/p(x)$
\footnote{
    \ In the next section we will use the notation 
    $r(a; b| c) = \frac{p(a,b|c)}{p(a|c)p(b|c)}$, 
    where with `$|$' we refer to conditioning on specific variables for all factors in the ratio definition. If necessary, multiple variables are comma separated, for example $r(a, b; c| d) = \frac{p(a, b, c|d)}{p(a, b|d)p(c|d)}$. Training conditional ratios is a straight-forward extension of \gls*{nre}.
},
which we can use to obtain weighted samples from the posterior $p(z | x) = r(z; x) p(z)$.
%
%
In order to improve the network's learning and maximise the simulator efficiency, \gls*{tmnre} concentrates in stages the regions in parameter space from which training examples are drawn based on a target observation. Hence, this truncation scheme restricts the prior distribution's support without modifying its shape, as opposed to other sequential methods that employ a posterior estimate as proposal distribution for generating simulations for the next round~\citep{Durkan_2018}.

\paragraph{Simulation model}
\label{par:model}
We consider here a simple Bayesian hierarchical source model,
\begin{equation}
    p(\bx, \vec \bs, \bvartheta) = p(\bx| \vec \bs) p(\bvartheta) \prod_{i=1}^N p(\bs_i|\bvartheta)\;,
    \label{eqn:model}
\end{equation}
where $\bx$ is the observed sky map, $\bs_i \equiv (F_i, \Omega_i)$ denotes the flux $F_i$ and position $\Omega_i$ of point source $i$, and $\bvartheta \equiv \{N, \Sigma, h\}$ collects source population parameters, namely the number of sources $N$, and the parameters $\Sigma$ and $h$ that control the flux and spatial distributions respectively. Finally, we add instrumental effects to simulated maps. We give more details on the model in \cref{apx:model}. We show examples of our simulated maps in the top row of the right panel of~\cref{fig:data}.

\paragraph{Source detection}
For source detection we consider the following likelihood-to-evidence ratio
\begin{equation}
    { r_1(\Omega, F_{th}; \bx)} 
    \equiv \frac{
    p(\mathbb{I}_{\bx}(F \geq F_{th})=1, \Omega|\bx)
    }{
    p(\mathbb{I}_{\bx}(F \geq F_{th})=1, \Omega)
    }\;.
    \label{eqn:rFth}
\end{equation}

Here, the denominator corresponds to the prior probability of having a source at position $\Omega$ with a flux $F\geq F_{th}$ that exceeds some threshold flux $F_{th}$.  The numerator is the corresponding posterior. 
We model the source detection ratio estimator in~\cref{eqn:rFth} as an image-to-image neural network that solves a binary classification problem in each image pixel. 
For simulated data, we call a simulated source $\bs_i$ `detected' when there is a corresponding compact region as function of $\Omega$ where the detection significance is above threshold, ${ r_1(\Omega, F_{th}; \bx)} > 5$. This effectively leads to a split between sources that are clearly identifiable and `sub-threshold' sources that are difficult to detect a individual instances.
Below, we assign the detection label $d_i = 1$ ($d_i = 0$) to detected (undetected) sources.

In order to characterize the split between detected and sub-threshold sources, we introduce a source-sensitivity function, $S(F, \Omega)$, which provides the probability that a source with flux $F$ and at position $\Omega$ would be detected by the ratio estimator in~\cref{eqn:rFth}.  This function can be estimated by training the ratio estimator
\begin{equation}
    { r_2(d;F, \Omega, \bx)}
    \equiv \frac{p(d| F, \Omega, \bx)}
    {p(d)}
    \label{eqn:rd}
\end{equation}
which is marginalised over all other sources and source parameters. By omitting the dependence on the map, $\bx$, which is then effectively marginalized, the ratio estimator can then simply be modeled as a $\mathbb{R}^3 \to \mathbb{R}$ multi-layer perceptron. The source-sensitivity function can then be estimated as 
\begin{equation}
S(F, \Omega) = \sigma\left(\log \left(\frac{p(d=1|F, \Omega)}{p(d=0|F, \Omega)}\right)\right),
\end{equation}
where we have introduced the sigmoid function $\sigma(y) \equiv 1 / (1 + e^{-y})$.

We can make the concept of source detection part of our model as follows.  In a random realization, each source $i$ will be either detected, $d_i= 1$ (with probability $S(F_i, \Omega_i)$), or not detected, $d_i=0$ (with probability $1-S(F_i, \Omega_i)$).  To keep notation simple, we omit $d_i$ and instead group detected and non-detected (sub-treshold) sources together in vectors with corresponding subscripts, and write $\vec \bs \equiv (\vec \bs_{det}, \vec \bs_{sub})$.
Taking into consideration this split, our model becomes
\begin{equation}
    p(\bx, \vec \bs_{det}, \vec \bs_{sub}, \bvartheta) = p(\bx| \vec \bs_{det}, \vec \bs_{sub})
    p(\vec \bs_{det}|\bvartheta) 
    p(\vec \bs_{sub}|\bvartheta) 
    p(\bvartheta)\;.
    \label{eqn:splitmodel}
\end{equation}

Importantly, although $p(\vec \bs_{det/sub}|\bvartheta)$ depends on the sensitivity function $S(F_i, \Omega_i)$, the distribution of all sources $p(\vec \bs|\bvartheta)$ is the same as in~\cref{eqn:model}.

\paragraph{Detection as truncation}
Truncating source priors in~\cref{eqn:model} is difficult due to the the label switching problem ($\bx$ is invariant under relabeling sources).  
However, once specific sources are detected, they can be labeled and ordered arbitrarily.  
Let us assume that $N_{det}$ sources were detected by the ratio estimator in~\cref{eqn:rFth} in the regions $\mathcal{R}_i$ with $i = 1, \dots, N_{det}$ for a given observation of reference $\bx_o$.  Our ansatz for the indicator function, which selects a specific prior region, is
\begin{equation}
    \mathbb{I}_{\bx_o}(\vec \bs_{det}) = \prod_{i=1}^{N_{det}}
    \mathbb{I}_{\bx_o}(\Omega_i \in \mathcal{R}_i)
    \mathbb{I}_{\bx_o}(F_i \geq F_{th})\;.
    \label{eqn:I}
\end{equation}
Our truncation strategy is now to focus on the parameter space where $\mathbb{I}_{\bx_o}(\vec \bs_{det})=1$ for our data of interest $\bx_o$.
We can then write our truncated model as 
\begin{equation}
\centering
\scalebox{0.949}{$
   p(\bx, \vec \bs_{det}, \vec \bs_{sub}, \bvartheta,
   \mathbb{I}_{\bx_o}(\vec \bs_{det}))
   = p(\bx| \vec \bs_{det}, \vec \bs_{sub})
   p(\vec \bs_{det}|\bvartheta,\mathbb{I}_{\bx_o}(\vec \bs_{det}))
   p(\vec \bs_{sub}|\bvartheta) 
   p(\bvartheta)
   p(\mathbb{I}_{\bx_o}(\vec \bs_{det})|\bvartheta).
   $}
    \label{eqn:truncatedmodel}
\end{equation}

\begin{figure}
    \centering
    \includegraphics[height=4.5cm, keepaspectratio, clip]{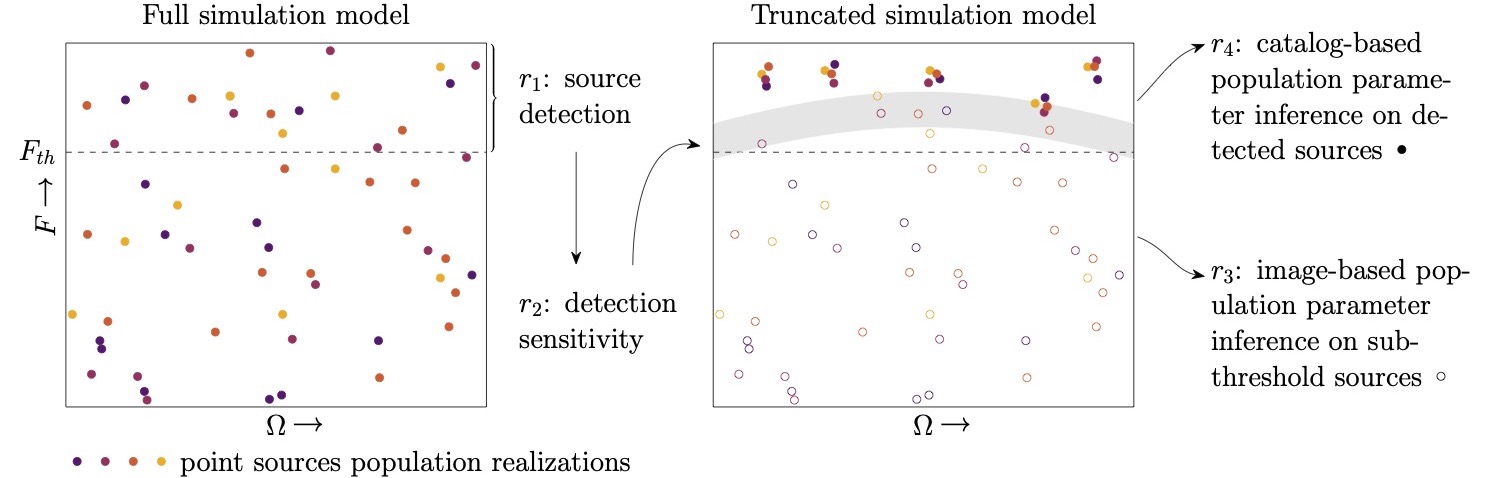}
    \caption{Illustration of our inference framework (see~\cref{sec:method} for details). \textit{Left panel}: A source detection network $r_1$ and the corresponding sensitivity network $r_2$ are trained based on the full simulation model. \textit{Right panel}: Bright sources are constrained in the truncated simulation model, while sub-threshold sources vary freely.  Two inference networks are trained to capture information from sub-threshold sources ($r_3$) and detected sources ($r_4$).
    }
    \label{fig:graphic}
\end{figure}

\paragraph{Population parameters inference with truncation}
Given some observation $\bx_o$, we want to estimate the posterior $p(\bvartheta|\bx_o)$. Since the truncation affects parameters $\vec \bs_{det}$ whose prior depends on population parameters $\bvartheta$, the truncation volume is not a constant factor, and the procedure requires extra care. We can estimate the posterior $p(\bvartheta|\bx)$ by considering the ratio
\begin{equation}
\centering
    r(\bvartheta; \bx) =
    \cfrac{p(\bvartheta|\bx)}{p(\bvartheta)}
    \simeq \cfrac{p(\bvartheta|\bx, \mathbb{I}_{\bx_o}(\vec\bs_{det})=1)}{p(\bvartheta)}
    = 
    \underbrace{\frac{p(\bvartheta|\bx, \mathbb{I}_{\bx_o}(\vec\bs_{det})=1)}{p(\bvartheta|\mathbb{I}_{\bx_o}(\vec\bs_{det})=1)}}_{\equiv { r_3(\bvartheta; \bx|\mathbb{I}_{\bx_o}(\vec\bs_{det})=1)}}
    \underbrace{\frac{p(\bvartheta|\mathbb{I}_{\bx_o}(\vec\bs_{det})=1)}{p(\bvartheta)}}_{\equiv
    r(\bvartheta; \mathbb{I}_{\bx_o}(\vec\bs_{det})=1)}\;.
    \label{eqn:posterior}
\end{equation}

The second step in~\cref{eqn:posterior} corresponds to the truncation approximation.
It is exact (in the sense of leaving $p(\bvartheta|\bx_o)$ unaffected) in the limit where $p(\bx_o|\mathbb{I}_{\bx_o}(\vec\bs_{det})=0) \to 0$.
Training that ratio estimator directly with \gls*{tmnre} would be challenging since $\mathbb{I}_{\bx_o}(\vec\bs_{det})=1$ has very small support in the training data.  Instead, we split the ratio into two computationally feasible ratios (this is in spirit similar to the telescoping ratio estimation approach presented in~\citet{telescoping}).

The first ratio, $ r_3(\bvartheta; \bx|\mathbb{I}_{\bx_o}(\vec\bs_{det})=1)$, can be estimated by training a peak-count network~\citep{Cholakkal_2019, Ranjan_2021, Kilic_2021} on targeted data, that is truncated to $\mathbb{I}_{\bx_o}(\vec\bs_{det})=1$. The second ratio can be estimated as
\begin{multline}
    r(\bvartheta; \mathbb{I}_{\bx_o}(\vec\bs_{det})=1)
    = \frac{p(\mathbb{I}_{\bx_o}(\vec\bs_{det})=1|\bvartheta)}{p(\mathbb{I}_{\bx_o}(\vec\bs_{det})=1)}\\
    = \int d\vec \bs_{det} \frac{p(\vec \bs_{det}|\bvartheta)} {p(\vec\bs_{det})}
    \frac{p(\vec\bs_{det})\mathbb{I}_{\bx_o}(\vec \bs_{det})}{p(\mathbb{I}_{\bx_o}(\vec\bs_{det})=1)}
    = \int d\vec \bs_{det} 
    \underbrace{\frac{p(\vec \bs_{det}|\bvartheta)} {p(\vec\bs_{det})}}_{
    \equiv {r_4(\vec\bs_{det};\bvartheta)}}
    p(\vec \bs_{det}|\mathbb{I}_{\bx_o}(\vec\bs_{det})= 1) \; ,
    \label{eqn:average}
\end{multline}
so it can be estimated by training a network on detected sources lists on un-truncated training data.
In practice, we generate weighted samples from the full posterior $p(\bvartheta|\bx)$ by sampling $\bvartheta, \vec \bs_{det} \sim p(\bvartheta)p(\vec \bs_{det}|\mathbb{I}_{\bx_o}(\vec\bs_{det})=1)$
with weights
$w = { r_3(\bvartheta; \bx|\mathbb{I}_{\bx_o}(\vec\bs_{det})=1)}\cdot
r(\bvartheta; \mathbb{I}_{\bx_o}(\vec\bs_{det})=1)
$.  

In the process, we trained four ratio estimation networks 
that are directly connected with traditional source analysis pipeline components: $ r_1(\Omega, F_{th}; \bx)$ performs source detection; $r_2(d; F, \Omega, \bx)$ is the source sensitivity function; $ r_3(\bvartheta; \bx|\mathbb{I}_{\bx_o}(\vec\bs_{det})=1)$ constraints $\bvartheta$ based on sub-threshold sources (because detected sources are assumed to be fixed in the parameter space where $\mathbb{I}_{\bx_o}(\vec \bs_{det})=1$); and $r_4(\vec\bs_{det}; \bvartheta)$ constraints $\bvartheta$ using the detected sources catalog.
We show a schematic overview of the inference framework used in this work in~\cref{fig:graphic}.

\section{Results}
\label{sec:results}

\begin{figure}
    \centering
    \includegraphics[width=\linewidth,clip]{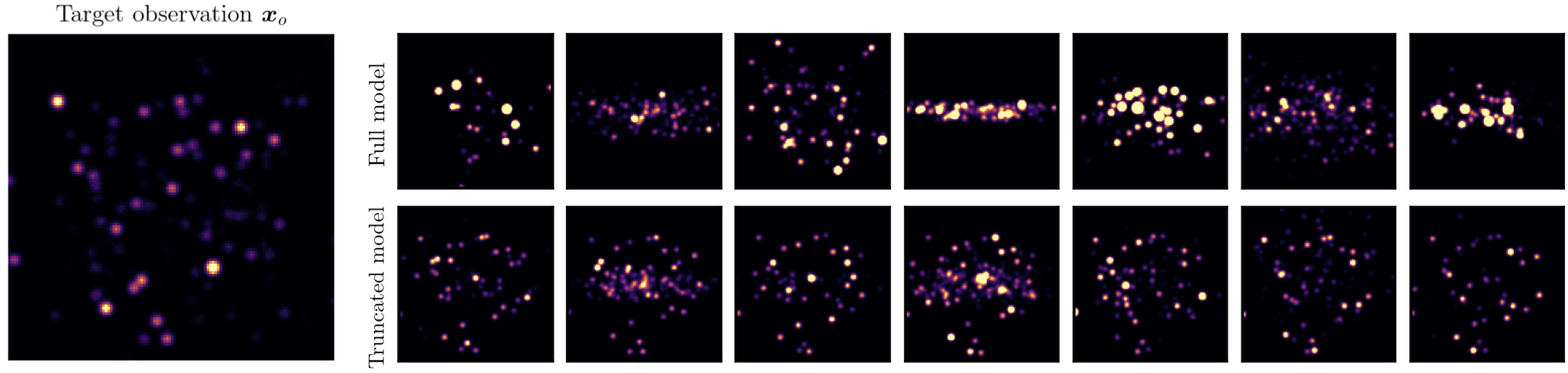}
    \caption{
      \textit{Left:} Observation of reference $\bx_o$. 
      \textit{Right:} In the first row we show samples from our full simulation model. In the second row we show targeted samples from the truncated model. Targeted data are visually more close to $\bx_o$, the main dissimilarities are due to different sub-threshold sources and instrumental effects realizations.
    }
    \label{fig:data}
\end{figure}

\begin{wrapfigure}[24]{R}{0.5\textwidth}
    \vspace{-35pt}
    \includegraphics[width=6.9cm]{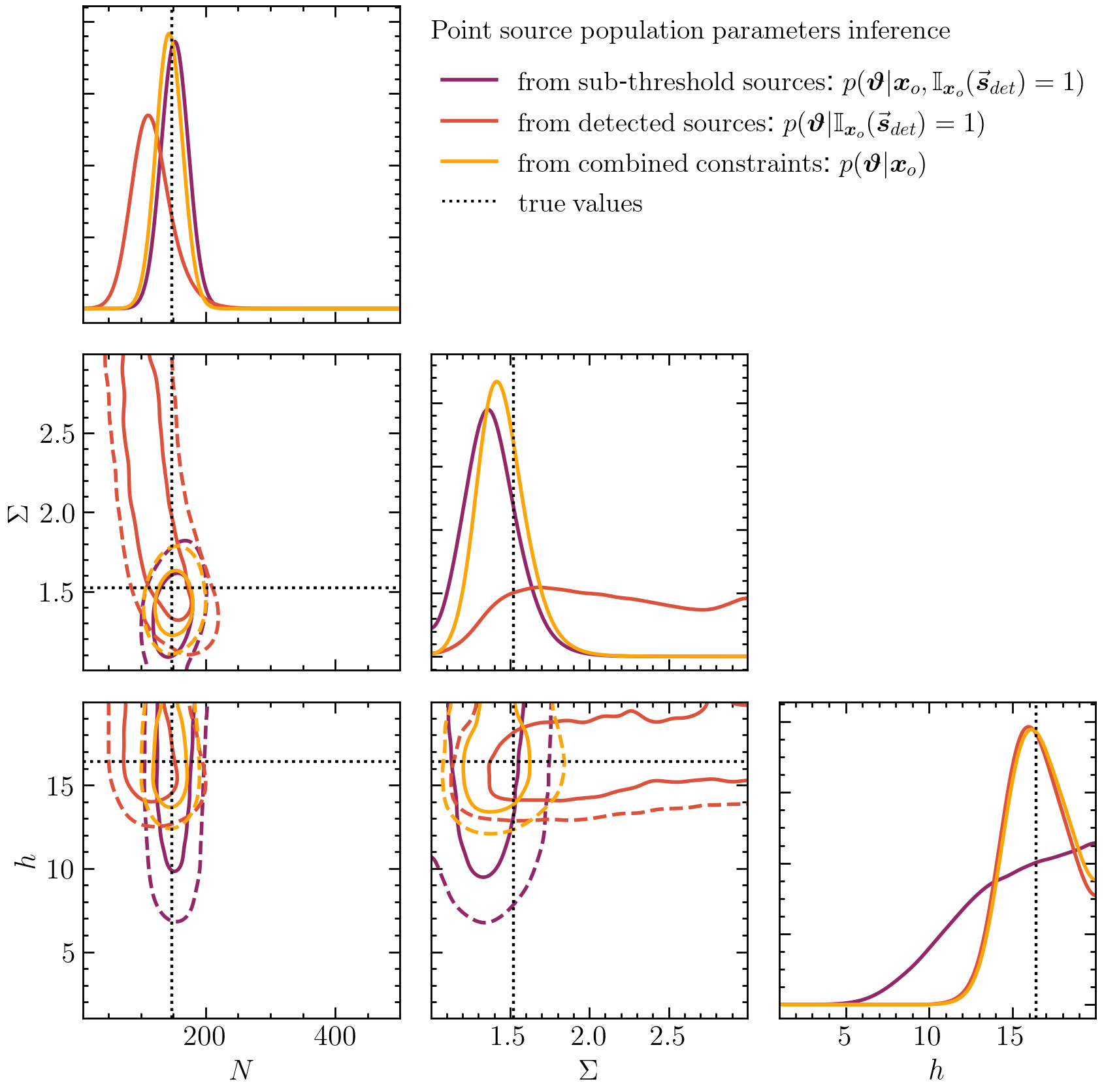}
    \caption{Marginal posteriors inferred by TMNRE on target observation $\bx_o$ for population parameters $\bvartheta$. We show constraints from sub-threshold sources (violet), detected sources (orange), and combined results (yellow). In the 2D marginal posterior we indicate the $68\%$ and $95\%$ credible regions with solid and dashed lines respectively. For more details see~\cref{sec:results}.
}
\label{fig:results}
\end{wrapfigure}

We apply the proposed methodology to the target observation $\bx_o$ shown in~\cref{fig:data}.  
We first train the source detection ratio estimator in~\cref{eqn:rFth} on data simulated from the full model shown in~\cref{eqn:model}, and then apply it to $\bx_o$ to obtain a detection map. From the detection map we derive a catalog of detected sources $\vec \bs_{det}$, and define a truncated parameter space of interest, where $\mathbb{I}_{\bx_o}(\vec \bs_{det})=1$.
In order to make source detection part of our model, we train the sensitivity ratio estimator in~\cref{eqn:rd} to estimate the sensitivity function $S(F, \Omega)$. 
We then generate targeted training data from our truncated simulation model in~\cref{eqn:truncatedmodel}. We show samples from the full model and the truncated one in~\cref{fig:data}. 

Finally, we train two inference networks to capture information regarding population parameters from sub-threshold and detected sources, as explained in~\cref{sec:method}.
We show the constraints on population parameters $\bvartheta$  from sub-threshold sources, detected ones, and their combination in~\cref{fig:results}. The different posteriors are consistent with each other, indicating that the proposed inference framework automatically accounts for detection biases. 
We see that different constraints are dominated by different neural networks, e.g. the number $N$ of point source is better constrained by sub-threshold ones, whereas the spatial distribution parameter $h$ by detected ones.
The weaker constraint on the flux parameter $\Sigma$ inferred from detected sources is due to the fact that in~\cref{eqn:average} we average over different detected sources realisations, always re-sampling the parameter $\Sigma$ (see~\cref{apx:model} for the hierarchical model details).

The four neural networks were trained on a NVIDIA GeForce RTX 3080 Ti GPU, the total computational time cost to obtain the results shown in~\cref{fig:results} is $\sim 2$ hours. We intend on describing in details the networks used for each task and choice of hyperparameters in future work.

\vspace{22pt}
\section{Conclusions}
\label{sec:conclusions}

We have introduced a novel method to self-consistently perform point sources detection and source population parameters inference using \gls*{tmnre}. With this approach, we can exploit information of detected as well as sub-threshold sources for population-level parameter inference. Detection biases are automatically accounted for in our approach.
Exemplary results of our approach are shown in~\cref{fig:results}, where we show inference results on source population parameters from both detected point sources and sub-threshold sources separately, as well as their combination. 
Since the proposed method is essentially a specific implementation of \gls*{tmnre}, we expect that it inherits its positive properties in terms of simulation-efficiency and scalability~\cite{miller_2021}.
A possible shortcoming of this approach is that multiple neural networks need to be trained self-consistently, which on the other hand have a clear interpretation in terms of traditional analysis pipeline components.
A potential application beyond those directly intended is to detectable and sub-threshold substructures in strong gravitational lenses.



\vspace{20pt}
\paragraph{Broader Impact}
\label{par:impact}
This work is focusing on the analysis of astronomical surveys that contain point sources population via \gls*{tmnre}. Variants of the presented approach could find application in other areas of the physical sciences. We do not expect any negative societal impact of the presented methods. However, we recommend the usual caution in inferring scientific conclusions based on a complex methodology.


\section*{Acknowledgments and Disclosure of Funding}

This work is part of a project that has received funding from the European Research Council (ERC) under the European Union’s Horizon 2020 research and innovation program (Grant agreement No. 864035 -- Undark).

We acknowledge the use of the \texttt{python} \citep{python} modules, \texttt{matplotlib} \citep{matplotlib}, \texttt{numpy} \citep{numpy},  \texttt{scipy} \citep{scipy}, \texttt{PyTorch} \citep{pytorch}, \texttt{tqdm} \citep{tqdm}, and \texttt{jupyter} \citep{jupyter}.


\bibliographystyle{unsrtnat}
\bibliography{bibliography}

\clearpage


\begin{appendix}

\section{Appendix: Simulation model}
\label{apx:model}

We describe in more detail the Bayesian hierarchical point source model adopted in this work. To generate an observation, first, we sample point sources population parameters $\bvartheta \equiv \{N, \Sigma, h\}$ from their priors, given in~\cref{tab:model}. Then, for each point source, we draw its flux $F$ and position on the map $\Omega\equiv(l, b)$ from the priors given in~\cref{tab:model}. We then generate a $128 \times 128$ pixels map. To model instrumental effects, we add a \gls*{psf} with Gaussian kernel standard deviation $\epsilon=1.5$ and Poisson noise, obtaining the final simulated map $\bx$. 

\begin{table}
  \caption{Point source simulation model parameters and priors.}
  \label{tab:model}
  \centering
  \begin{tabular}{l l l }
    \toprule
        Parameter & & Prior \\
    \midrule
        Population parameters & & \\
            \quad number of point sources & $N$ & $\mathcal{U}(10, 500)$ \\
            \quad flux distribution parameter & $\Sigma$ & $\mathcal{U}(1, 3)$ \\
            \quad spatial distribution parameter & $h$ & $\mathcal{U}(1, 20)$ \\
    \midrule
        Point source parameters & & \\
            \quad flux & $F$ & $\mathcal{\log N}(1, \Sigma)$ \\
            \quad position & $\Omega\equiv(l, b)$ & ($\mathcal{N}(0, 20)$, $\mathcal{N}(0, h)$) \\
    \bottomrule
  \end{tabular}
\end{table}

\end{appendix}

\clearpage

\end{document}